\documentclass{article}

\usepackage{spconf}
\usepackage{amsmath}
\usepackage{times}
\usepackage{amssymb}
\usepackage{multirow}
\usepackage{array}
\usepackage{url}

\usepackage{epsfig}
\usepackage{graphicx}
\graphicspath{{figures/}}

\def\x{{\mathbf x}}

\def\etal{\textit{et al.}}

\def\ie{\textit{i.e.}}

\def\eg{\textit{e.g.}}

\begin{document}
\ninept
 
\title{A proposal project for a blind image quality assessment by learning distortions from the full reference image quality assessments}
 
\name{St\'efane Paris}
\address{Universit\'e de Lorraine, Loria, UMR 7503, Vand{\oe}uvre-L\`es-Nancy}

\maketitle

\begin{abstract}
This short paper presents a perspective plan to build a null reference image quality assessment.
Its main goal is to deliver both the objective score and the distortion map for a given distorted image without the knowledge of its reference image.
\end{abstract}


\section{Introduction}

The Holy Grail of image quality assessment (IQA) is to be blind, \ie, to provide an IQA with null reference (NR-IQA). At first, there was full reference image quality assessments (FR-IQA) whose inputs are the original and distorted images. This was soon followed by the reduced reference version (RR-IQA) which proposed to reduce as much as possible the original data needed to assess the distorted image. Nowadays, these first assessment algorithms are mature although improvements can still be obtained, especially in RR-IQA. This first category of assessments with complete or reduced reference is of two parts\,: the processes delivering an objective score (O-Score) and those delivering both a O-Score and a distortion map. Among the former we can quote~\cite{zhai12} and~\cite{wang11b} as being in the top ten. From the latter we can quote the well-known SSIM~\cite{wang04,wang11a}. The fact that an assessment process can or cannot deliver a distortion map is of importance; the paper written by C T. Vu \etal~\cite{
vu12} illustrates this for the special case of sharpness perception. In other words, the fact that the global O-Score of a distorted image is strongly correlated to the subjective score of this image does not provide the localizations of the distortions.

Thus, the current project of FR- and NR-IQA concentrates on algorithms delivering both an objective score and a distortion map. The FR-IQA was already presented in \cite{paris10a} and is briefly described in section~\ref{sec:weqa}. But any FR-IQA, \eg~SSIM, can be used in place.

\begin{figure}[tbh]
\centering\includegraphics[width=0.4\textwidth]{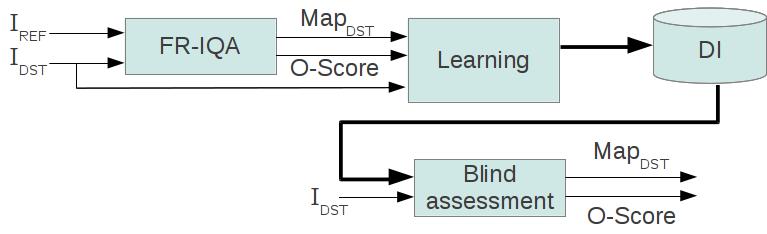}
\caption{
	The two-stage schema of the presented blind process; the learning stage is based on the
	FR-IQA process which works offline; the second stage is online and should deliver both an
	objective score and a distortion map with the help of the database learned from the results of the
	FR process.
} 
\label{fig:blind}
\end{figure}

This short paper proposes a model of the main stages of the NR-IQA which learns to identify the distortion localizations from the results of the FR-IQA. Figure~\ref{fig:blind} shows the global diagram of our current study on blind assessments.
The next section explains the desired two-stage process.
Section~\ref{sec:Two-stage-process} details some aspects the process should have. Section~\ref{sec:weqa} explains the FR-IQA. This paper is concluded in section~\ref{sec:Perspective-plan}

\section{Global scheme}

As shown in figure~\ref{fig:blind} the learning stage is based on the FR-IQA
WEQA which delivers both an O-Score and a distortion map~\cite{paris10a,paris11}.
The O-Scores were statistically studied in~\cite{paris11}.
The Spearman's rank-order and Pearson's~$\chi{}^{2}$ correlation coefficients show good performances relative to the statistics provided by SSIM and VIF~\cite{lin11}.
Moreover the distortion map of WEQA is at least as precise as the one delivered by SSIM~\cite{wang04}.

With the help of the distortion maps delivered by of WEQA, a distortion database
such as LIVE~\cite{livedb2} or TID2008~\cite{tid} can be learned.
As the learning phase is pixel-based, two questions arise\,:

\begin{itemize}
	\item Which pixels to learn and how to characterize them\,?
	\item How to classify these pixels and what do the clusters mean\,?
\end{itemize}

\section{Two-stage process} 
\label{sec:Two-stage-process}

At the learning phase, WEQA estimates a distortion map for each distorted image providing its reference image.
From each distorted image and its distortion map, each pixel is assigned a vector descriptor and a level of distortion which allow the learning process to cluster the pixels.
Using WEQA, the characterization can be the \emph{wave-vectors} and \emph{color-vectors}.
We can hypothesize that the neighbourhood of each pixel should be involved in the clustering.
Currently, we are studying the codebook to be generated from these wave- and color-vectors.
We specially focus on information that the distortion maps provide.

The content-based image retrieval (CBIR) community proposes several effective algorithms for
indexing and retrieving image in large image databases. The well-known SIFT~\cite{lowe04} uses an
optimized kd-tree to index the scale-invariant keypoints. But, F. Moosmann \etal~have shown
in~\cite{moosmann06} that indexing structures like kd-tree are not efficient with huge image
databases. Following the study of Geurts \etal~on extremely randomized trees\cite{geurts06}, they
proposed a learning scheme based on support vector machine (SVM). J. Shotton
\etal~\cite{shotton08} extended the principle to semantic texton forests such that even the
characterization of pixels was randomized.

The current project of NR-IQA is inspired by these studies.
All the pixels of the distorted images can be learned through randomized forests.
The forests would provide the kernels of the SVMs to use during the blind phase.

\section{The full refrence image quality assessment}
\label{sec:weqa}

\begin{figure}[htb]
\centering\includegraphics[width=0.49\textwidth]{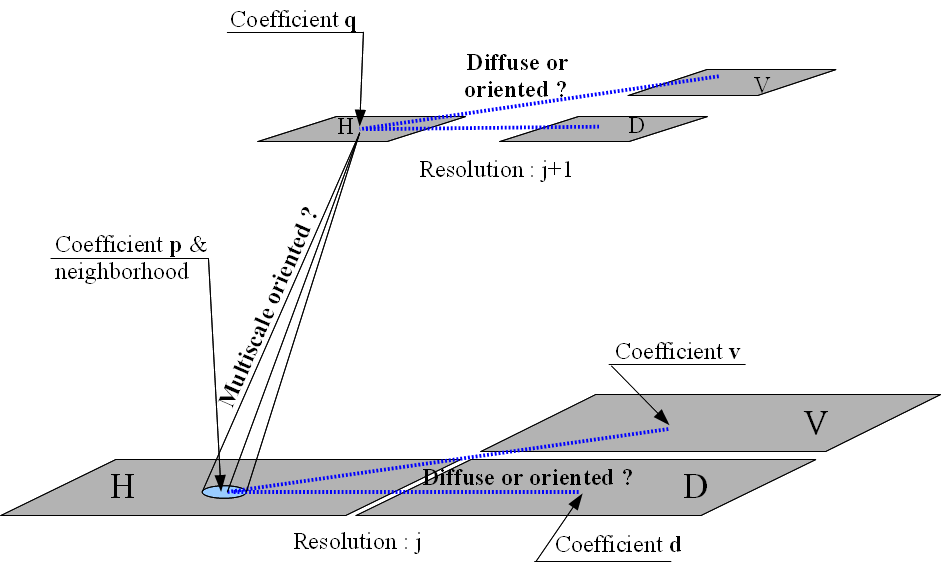}%
\caption{%
	This partial wavetree shows the relationship between coefficients at different scales and the relationship between coefficients at the same level of resolution. H, V and D denote the horizontal, vertical and diagonal sub-band respectively. The coefficient \textbf{q} is the strength of the horizontal direction at resolution j$+$1. It is related to coefficients \textbf{p} and its neighbourhood\,: the stronger the coefficients \textbf{p} and \textbf{q} are, the more reliable the horizontal direction is. Besides, if the coefficient \textbf{p} is stronger than its corresponding coefficients, \textbf{d} and \textbf{v} from diagonal and vertical subbands, the horizontal direction is reinforced. Otherwise, no specific orientation is observed.
}
\label{fig:wavetree}
\end{figure}

The full reference image quality assessment algorithm is wavelet domain and based on the generalized Euclidean distance.
A wavelet analysis is performed  on the reference and distorted images separately.
This multiscale representation of the images provides an oriented description at each pixel\,: the quantity of pixels sharing a coefficient grows with the level of resolution of the coefficient.
By this fact, each pixel iwhich embeds orientation information (see Fig. \ref{fig:wavetree}).
\label{sec:Two-stage-process}To estimate the level of distortion of each pixel, we used the anisotropic image Euclidean distance~\cite{wang05}\,:
$d^{2}_{\mbox{\textsc{weqa}}}(p) = \sum_{i,j=1}^M g_{i,j} \Delta_{p,i} \Delta_{p,j}$\,;
with $g_{i,j} = \exp{\left( -(i-j)^2\right)/2}$, $\Delta_{p,i} = (\Phi_{p,i} - \Phi^\prime_{p,i})$ and where $\mathbf{\Phi}_{p} = (\Phi_{p,1}, \ldots , \Phi_{p,M})^T$ and $\mathbf{\Phi}^\prime_{p}$ are the wave-vectors of the reference and distorted images respectively.
If $\mathbf{\Phi}_{p}$ and $\mathbf{\Phi}^\prime_{p}$ are of equal magnitude, $\|\Phi\|^2$, the  distance looks for similar coefficients.
If it is, the two wave-vectors are perceived closer than if they do not share any coefficient.
In other words, this distance brings closer the pixels with similar orientations and/or similar activities in their neighbourhood.
For more details on the FR-IQA and its performances see \cite{paris10a,paris11}.

\section{Perspective plan}
\label{sec:Perspective-plan}
 
Nowadays, no results are available. The current work focuses on the pixel selection and the different  characterizations of these pixels based on their wave- and color-vectors.
We search for a modelling of these descriptors in the context of the extremely randomized trees.

The model as presented here is available  for one kind of distortion.
If it gives significant results, we shall extend the distribution modelling of the learning process to bring together different kinds of distortions.

We hope our conjectures will be validated at least partly, \ie, when treating each kind of distortion independently.
The biggest challenge will be to deal with all the kinds of distortion together.
More information can be find at \url{https://sites.google.com/site/imagequalityassessment/}

{\small
\bibliographystyle{IEEEbib} 
\bibliography{/home/stefane/Bibliotheque/biblio}
}

\end{document}